\documentclass[prb,twocolumn,superscriptaddress,amssymb,longbibliography]{revtex4-2}
\usepackage[utf8]{inputenc}
\usepackage[T1]{fontenc}
\usepackage{graphicx}
\usepackage{dcolumn}
\usepackage{bm}
\usepackage{epsfig}
\usepackage{float} 
\usepackage{color}
\usepackage{tikz}
\usepackage{subcaption}
\usepackage{verbatim}
\usepackage{float}
\usepackage[normalem]{ulem}
\usepackage{multirow}
\usepackage[colorlinks=true,urlcolor=blue,linkcolor=blue,
            citecolor=blue]{hyperref}

\newcommand \beq{\begin{equation}}
\newcommand \eeq{\end{equation}}

\newcommand \exciting{\texttt{Exciting}}

\newcommand \kvec{{\bf k}}

\newcommand \qvec{{\bf q}}

\newcommand\rvec{{\bf r}}

\newcommand\Gvec{{\bf G}}

\newcommand{\rhot}{\tilde{\rho}}
\newcommand{\degree}{^\circ}

\newcommand{\alo}{$\alpha$-Al$_2$O$_3$}

\def\simge{\mathrel{%
       \rlap{\raise 0.511ex \hbox{$>$}}{\lower 0.511ex \hbox{$\sim$}}}}
\def\simle{\mathrel{
       \rlap{\raise 0.511ex \hbox{$<$}}{\lower 0.511ex \hbox{$\sim$}}}}
\def\beq {\begin{equation}}
\def\eeq {\end{equation}}

\def\w {\omega}
\def\bfq {\mathbf{q}}

\def\bfk {\mathbf{k}}

\begin{document}

\title{K-edge XANES of octahedral aluminum compounds: similarities and differences via the analysis of excitonic properties}

\newcommand{\lsi}{LSI, CNRS, CEA/DRF/IRAMIS, \'Ecole Polytechnique, Institut Polytechnique de Paris, F-91120 Palaiseau, France}
\newcommand{\etsf}{European Theoretical Spectroscopy Facility (ETSF)}
\newcommand{\soleil}{Synchrotron SOLEIL, L'Orme des Merisiers, Saint-Aubin, BP 48, F-91192 Gif-sur-Yvette, France}

\author{Newman Amoyaw}
\thanks{These two authors contributed equally.}
\affiliation{Department of Physics, School of Physical and Mathematical Sciences, College of Basic and Applied Sciences, University of Ghana, Legon, Accra, Ghana}

\author{Abezu Agegnehu}
\thanks{These two authors contributed equally.}
\affiliation{Applied Physics Department, Adama Science and Technology University, P.O. Box 1888, Adama, Ethiopia}

\author{Francesco Sottile}
\affiliation{\lsi}
\affiliation{\etsf}

\author{Matteo Gatti}
\affiliation{\lsi}
\affiliation{\etsf}
\affiliation{\soleil}

\author{M. Laura Urquiza}
\affiliation{\lsi}
\affiliation{\etsf}

\begin{abstract}

This study presents an \textit{ab initio} investigation of the XANES spectra at the aluminum K edge for three compounds:  Al$_2$O$_3$, AlF$_3$ and AlCl$_3$, where the Al atoms share the same oxidation state~(III) and are coordinated in an octahedral symmetry. 
The XANES spectra calculated within the independent-particle approximation reveal significant differences, including shifts in the spectrum onset, variations in the spectral shapes, and the presence of a pre-peak in the case of AlCl$_3$, all in correspondence with the behavior of the PDOS of the absorbing atom in the different materials.
The origin of the features stems from the specific band structure of each compound. 
When electron--hole interactions are taken into account through the solution of the Bethe-Salpeter equation, a series of dark and bright excitons with large binding energies and Frenkel character is obtained. The strong excitonic effects lead to the suppression of the pre-peak in AlCl$_3$ and further accentuate the differences among the three Al K-edge spectra.
\end{abstract}

\maketitle

\section{\label{sec:introduction} Introduction}

X-ray absorption spectroscopy, particularly x-ray absorption near edge structure (XANES), is a powerful tool for the structural characterization of materials. 
XANES provides high sensitivity to the local chemical environment of the absorbing atom and to the medium-range structural organization, yielding valuable information on symmetry~\cite{Dien_Bancroft_1995,Ildefonse_1998}, coordination number~\cite{Ildefonse_1998,Bokhoven_2001,Weigel_2008}, bond length~\cite{Bianconi_1985} and bond angle. Therefore, it serves as a distinctive fingerprint to identify and characterize constituent atoms within a material's structure.
XANES can be also used to infer changes in the oxidation state of the absorbing atom. An example is the K edge of Mn, where changes in the oxidation state of Mn from II to IV in a series of compounds yield different chemical shifts~\cite{Hanson_1949, Grush_1995}. Similarly, in a series of vanadium oxides the energy shifts of the absorption edge and the pre-edge have been observed to be linearly dependent on the valence of the absorbing vanadium atom~\cite{Wong_1984}.

The fine structures measured with XANES provide information on the local density of the empty electronic states at the site of the absorbing atom, in the presence of the core hole.
In particular, K-edge XANES spectra probe predominantly $s\rightarrow p$ dipole transitions, since monopole ($s \rightarrow s$) and quadrupole ($s\rightarrow d$) terms are normally orders of magnitude smaller. 
However, a pre-edge related to dipole-forbidden transitions is often present at the K edge of low-Z cations in minerals and oxides.
These pre-edge features are determined by multiple factors, including coordination number, local distortion, oxidation state, and the nature of ligands around the absorbing atom. 
For example, transition metal compounds with tetrahedral geometries usually exhibit stronger pre-peak intensities than those with octahedral geometries~\cite{Yamamoto_2008,Srivastava_1973,Jiang_2007}.  
The pre-edge features stem from the density of states the bottom of the conduction band, which
in these compounds is composed primarily by the transition metal $3d$ empty states. 
According to group theory, atomic $p-d$ mixing is forbidden under octahedral symmetry but allowed under tetrahedral symmetry. When absorbing atoms have a tetrahedral or distorted octahedral coordination, this results into an enhancement of the pre-peak intensity, where dipole transitions from the core $1s$ state to the $p-d$ hybridized orbitals at the bottom of the conduction band become possible~\cite{Westre_1997}. 
A similar behavior has been observed in aluminum compounds and minerals~\cite{Dien_Bancroft_1995, Ildefonse_1998, Cabaret_1996, Cabaret_2009, Bokhoven_2001, Weigel_2008}, where tetrahedral symmetry or local distortions in octahedral geometries enable atomic $s-p$ mixing.

Extensive research has been carried out to correlate K-edge spectral features, particularly in the pre-edge region, with the local chemical environment. 
However, understanding the spectral shapes in detail remains challenging. 
For instance, in aluminum compounds with octahedral geometries, six-fold coordinated aluminum exhibits a variety of XANES features. 
The interplay among  the various structural parameters (such as the number of Al sites and the distribution of interatomic distances) complicates the quantitative interpretation of XANES spectra~\cite{Dien_Bancroft_1995, Ildefonse_1998}. 
On the other hand, band-structure effects are also expected to strongly influence the absorption features. 
In addition, the interactions between the core holes and the electrons excited in the conduction states, also known as excitonic effects, can play a crucial role, leading to substantial shifts in the absorption edge and modifications in absorption intensities.
In this context, theoretical studies can provide important information for the interpretation of spectra by allowing the selective activation or deactivation of specific interactions at play in the materials. 

For excitations from deep core states, predicting near-edge structures requires appropriate theoretical methods, capable of reproducing spectral features and excitonic effects. 
In particular, the interaction between the excited electrons and core holes, highly localized and poorly screened, demands a high-level treatment of electron-hole correlation.
Common approaches for XANES spectra,
in the so-called \textit{core-hole} or \textit{final-state-rule} approximation~\cite{Rehr_2005}, typically evaluate a one-electron Fermi Golden rule, with a relaxed final state including a partial or a full core hole at a single atomic site within a supercell.  
Other methods based on multiplet ligand-field theory~\cite{deGroot_2001, deGroot_2008,deGroot_2021} are very efficient, but are not completely parameter free.
Alternatively, XANES spectra can be obtained within linear-response theory through the Bethe-Salpeter equation\cite{Strinati_1988,Onida_2002} (BSE) using an all-electron approach. 
Here, core electrons are treated explicitly within a muffin-tin region around the nuclear positions. 
The BSE is the state-of-the-art method for calculating optical excitations\cite{Onida_2002} and is also increasingly used for core-level excitation spectra in solids~\cite{Olovsson_2009, Olovsson_2009b, Olovsson_2011, Vinson_2011, Vinson_2012, Gilmore_2015, Gilmore_2021, Geondzhian_2015, Vorwerk_2017, Vorwerk_2019, Vorwerk_2020, Vorwerk_2022, Vinson_2022, Dashwood_2021, Unzog_2022}.

The present work aims to analyze changes in the K-edge XANES spectra of aluminum compounds and relate them to their crystallographic properties and chemical environments. We have carried out independent-particle and state-of-the-art BSE calculations to obtain XANES spectra at the aluminum K edge for Al$_2$O$_3$, AlF$_3$, and AlCl$_3$, all containing Al atoms with the oxidation state ${+3}$ and octahedral coordination. The results show the
limits of the atomic perspective. Distorsions from the ideal octahedral structure coupled with strong 
excitonic effects make the {\it ab initio} results crucially needed for a quantitative description and analysis 
of the aluminum compounds.

The manuscript is organized as follows. In Sec.~\ref{sec:method} we describe the theory for the calculation of the XANES spectra through the BSE and independent-particle approximation (IPA), together with the computational details. In Sec.~\ref{sec:results} we present the results for Al$_2$O$_3$, AlF$_3$, and AlCl$_3$.
Finally, in Sec.~\ref{sec:conclusions} we draw conclusions and perspectives of the present work.

\section{Methods \label{sec:method}}

\subsection{XANES spectrum from the BSE \label{ssec:bse}}

The XANES spectra are calculated by solving the Bethe-Salpeter equation (BSE), which gives the density-density
response function from the solution of a Dyson-like equation for a two-particle correlation function\cite{Strinati_1988}. 
Applying the GW approximation~\cite{Hedin_1965} and using a statically screened Coulomb interaction $W$, the BSE can be expressed as an effective two-particle Schr\"odinger equation~\cite{Onida_2002}: 
\begin{equation}
	H^{\rm exc} A_{\lambda} = E_{\lambda} A_{\lambda},
	\label{eq:bse1}
\end{equation}
where $H^{\rm exc}$ is the excitonic Hamiltonian and $E_{\lambda}$ are the excitation energies.  
The excitonic Hamiltonian is given by the sum of the three following terms: 
\begin{equation}
	H^{\rm exc} = H^{ipa} + 2H^{x} - H^{c}
	\label{eq:bse}
\end{equation}
which are matrices in the basis of transitions that, in the Tamm-Dancoff approximation, are 
between occupied core states 
and unoccupied  
Kohn-Sham orbitals  $\varphi_{\mu\kvec}\rightarrow\varphi_{c\kvec}$. 
More specifically,  
those matrix elements are: 
\begin{equation}
H^{ipa}_{c\mu\kvec,c'\mu'\kvec'} = E_{{\mu}c{\bf k}} \delta_{\mu \mu'}\delta_{cc'}\delta_{{\bf k}{\bf k}'} 
\end{equation}
\begin{equation}
H^{x}_{c\mu\kvec,c'\mu'\kvec'} = \int \varphi^{*}_{c\kvec}(\rvec) \varphi_{\mu\kvec}(\rvec) \bar{v}_c (\rvec,\rvec') \varphi_{c'\kvec'}(\rvec') \varphi^{*}_{\mu'\kvec'}(\rvec') d\rvec d\rvec' 
\end{equation}
\begin{equation}
H^{c}_{c\mu\kvec,c'\mu'\kvec'} = \int \varphi^{*}_{c\kvec}(\rvec) \varphi_{c'\kvec'}(\rvec) W(\rvec,\rvec') \ \varphi_{\mu\kvec}(\rvec') \varphi^{*}_{\mu'\kvec'}(\rvec') d\rvec d\rvec'.
\end{equation}

Here  $E_{{\mu}c{\bf k}}=E_{c\bfk}-E_{{\mu}\bfk}$ are the interband  transition energies,  $\bar v_c$ is the bare Coulomb interaction, calculated without its macroscopic component (i.e., the component ${\bf G}=0$ in reciprocal space is set to $0$), and the statically screened Coulomb interaction $W=\epsilon^{-1}v_c$ is obtained within the random-phase approximation (RPA) for the inverse  dielectric function $\epsilon^{-1}$. 
The first term in Eq. (\ref{eq:bse}), $H^{ipa}$, represents the independent-particle transitions. 
The repulsive term, $H^{x}$, is the exchange electron-hole interaction and is responsible for the local-field  effects (LFE)~\cite{Wiser_1963,Adler_1962}. 
LFE are important in inhomogeneous systems as they reflect the inhomogeneities in the induced microscopic Hartree potential that counteracts the external perturbation.
Finally, the attractive term, $-H^{c}$, describes the electron-hole attraction.

The XANES spectrum can be obtained from the imaginary part of the macroscopic dielectric function $\textrm{Im}\epsilon_M(\w)$ in the long-wavelength limit $\bfq\to0$, which 
can be calculated from of eigenvectors $A_\lambda$ and eigenvalues $E_\lambda$ of the BSE Hamiltonian, Eq. (\ref{eq:bse1}), as:
\begin{equation}
\textrm{Im}\epsilon_M(\w) =  \lim_{\qvec\to 0}\frac{8\pi^2}{\Omega q^2} \sum_\lambda \left|\sum_{\mu c\kvec}  A_\lambda^{\mu c\kvec} \tilde{\rho}_{\mu c\kvec}(\qvec) \right|^2 \delta(\w- E_\lambda) ,
\label{eq:spectrumBSE}
\end{equation}
where  $\Omega$ is the crystal volume and $\rhot_{\mu c\kvec}$ are the 
oscillator strengths: 
\begin{equation}
\rhot_{\mu c\kvec} (\qvec) = 
\int \varphi^*_{\mu\kvec-\qvec}(\rvec) e^{-i\qvec\rvec} \varphi_{c\kvec}(\rvec) d\rvec.
\label{eq:rhotw}
\end{equation}
Each excitonic peak in the spectrum, located at energy $E_\lambda$,
has an intensity given by the absolute square modulus in Eq. (\ref{eq:spectrumBSE}). If the intensity is negligibly small, the exciton is said to be dark, and bright otherwise.

In the independent-particle approximation, one could rewrite Eq.~(\ref{eq:spectrumBSE}) simply as:
\begin{equation}
	\textrm{Im}\epsilon_M(\w) =  \lim_{\qvec\to 0}\frac{8\pi^2}{\Omega q^2} \left|\sum_{\mu c\kvec}  \tilde{\rho}_{\mu c\kvec}(\qvec) \right|^2 \delta(\w- E_{vc\kvec} ) ,
	\label{eq:spectrumIPA}
\end{equation}
since the excitonic eigenvectors $A_\lambda$, which in Eq.~(\ref{eq:spectrumBSE}) mix independent-particle transitions, become $\delta_{\mu\mu'}\delta_{cc'}\delta_{{\bf k}{\bf k}'}$ ($H^{ipa}$ is a diagonal matrix).
If, additionally, the oscillator strengths 
$\tilde{\rho}_{\mu c\kvec}$ are approximately constant for $s{\to}p$ transitions (and zero for all other transitions), then the K-edge XANES spectrum becomes proportional to the $p$ component of the projected density of unoccupied  states (PDOS) of the absorbing atom.

\subsection{Computational details \label{ssec:details}}

The calculations were performed within an all-electron full-potential  linearized augmented plane-wave (FP-LAPW) method, as implemented in {\exciting} code~\cite{Gulans_2014}. 
Kohn-Sham density functional theory (KS-DFT) \cite{HK1964} calculations have been done within the local density approximation (LDA)~\cite{Kohn_1965} for the exchange correlation functional. 
We adopted the experimental  crystal structure of the three compounds. $\alpha-$Al$_2$O$_3$ has a rhombohedral structure with lattice parameters~\cite{Newnham_1962} $a=5.128$~{\AA} and $\alpha=55.287 \degree$. $\alpha-$AlF$_3$, also with rhombohedral unit cell, has parameters~\cite{Daniel_1990} $a=5.0314$~{\AA} and $\alpha=58.6772\degree$. Finally, AlCl$_3$ has a monoclinic unit cell with parameters~\cite{Troyanov_1992} $a=5.914$, $b=10.234$, $c=6.148$ and $\alpha=108.25\degree$.
The convergency of the ground-state density was achieved using a plane-wave cutoff, $R_{MT} |\Gvec+\kvec|_{max}=10$  for AlF$_3$ and $R_{MT} |\Gvec+\kvec|_{max}=9$ for Al$_2$O$_3$ and AlCl$_3$. Here, $R_{MT}$ is the muffin-tin radius of the anion, O (1.45~Bohr), Cl (1.45~Bohr) and F (1.39~Bohr), which is always smaller than  $R_{MT}=2$ Bohr for Al.

The RPA screening was calculated considering 100 conduction bands and a cutoff in the matrix size given by $|\Gvec+\qvec|_{max}=4, 7$ and $6$ $a_0^{-1}$ for Al$_2$O$_3$, AlF$_3$ and, AlCl$_3$, respectively, maintaining the same cutoff for the wavefunctions as for the ground state.

The BSE calculations for the XANES spectra are performed using a shifted
$\kvec$-point grid (i.e., not containing high-symmetry $\kvec$ points), which allows for a quicker convergence of the spectra\cite{Benedict_1998}. The adopted $\kvec$-point grids are $8\times8\times8$ for  Al$_2$O$_3$, $10\times10\times10$ for AlF$_3$ and $6\times6\times6$ for AlCl$_3$ with a $(0.05,~0.15,~0.25)$ shift. In all three cases, we observed slow convergence with respect to the number of empty bands. To achieve accurate results, we had to include 60, 50, and 100 empty bands for Al$_2$O$_3$, AlF$_3$, and AlCl$_3$, respectively.
Finally, the XANES spectra have been plotted using a Lorentzian broadening of 0.1 eV.

\section{Results and discussion \label{sec:results}}

\subsection{6-fold coordinated aluminum compounds \label{ssec:compounds}}

\begin{figure*}[t]
	\includegraphics[width=0.3\textwidth]{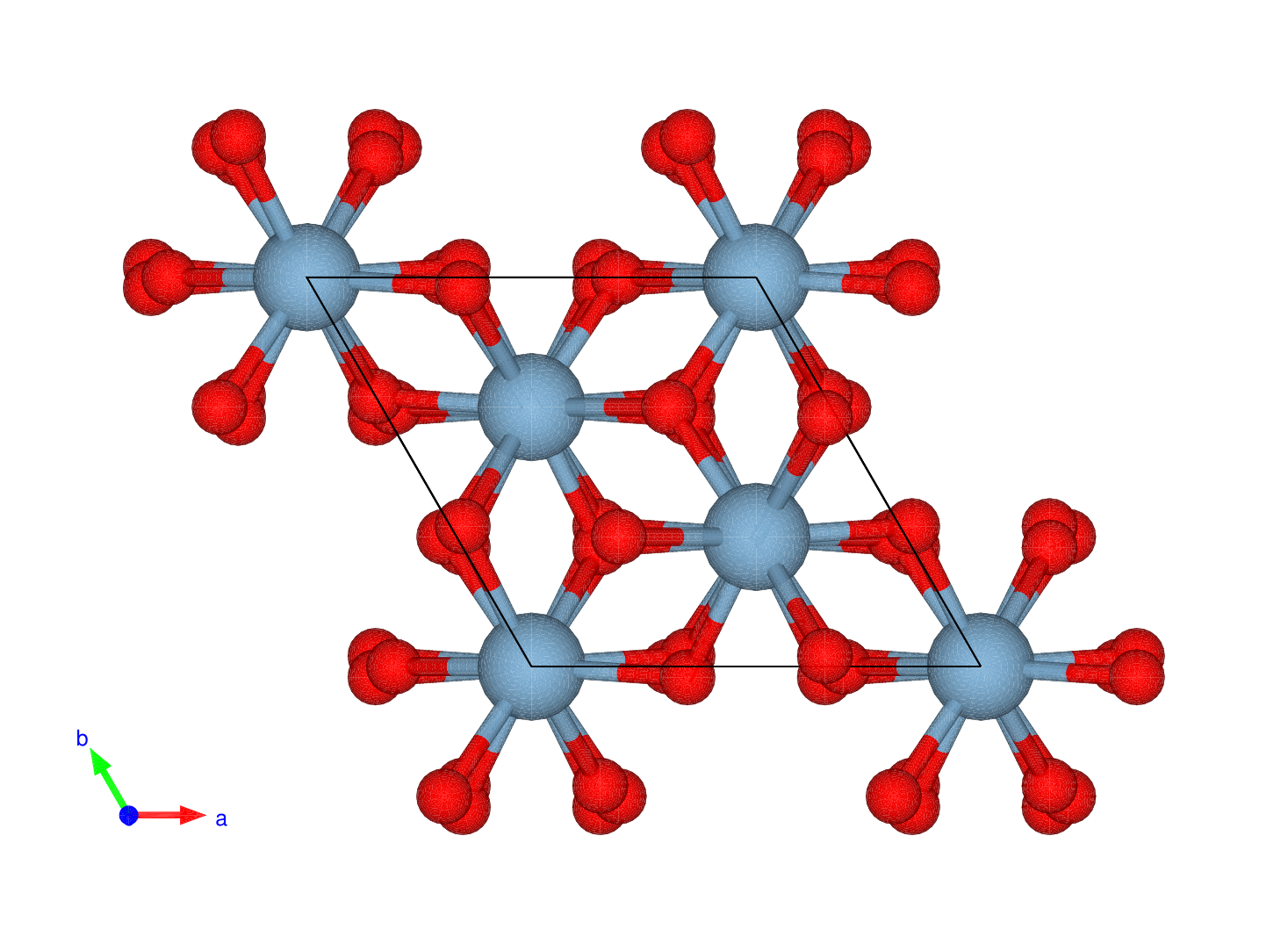}       
	\includegraphics[width=0.3\textwidth]{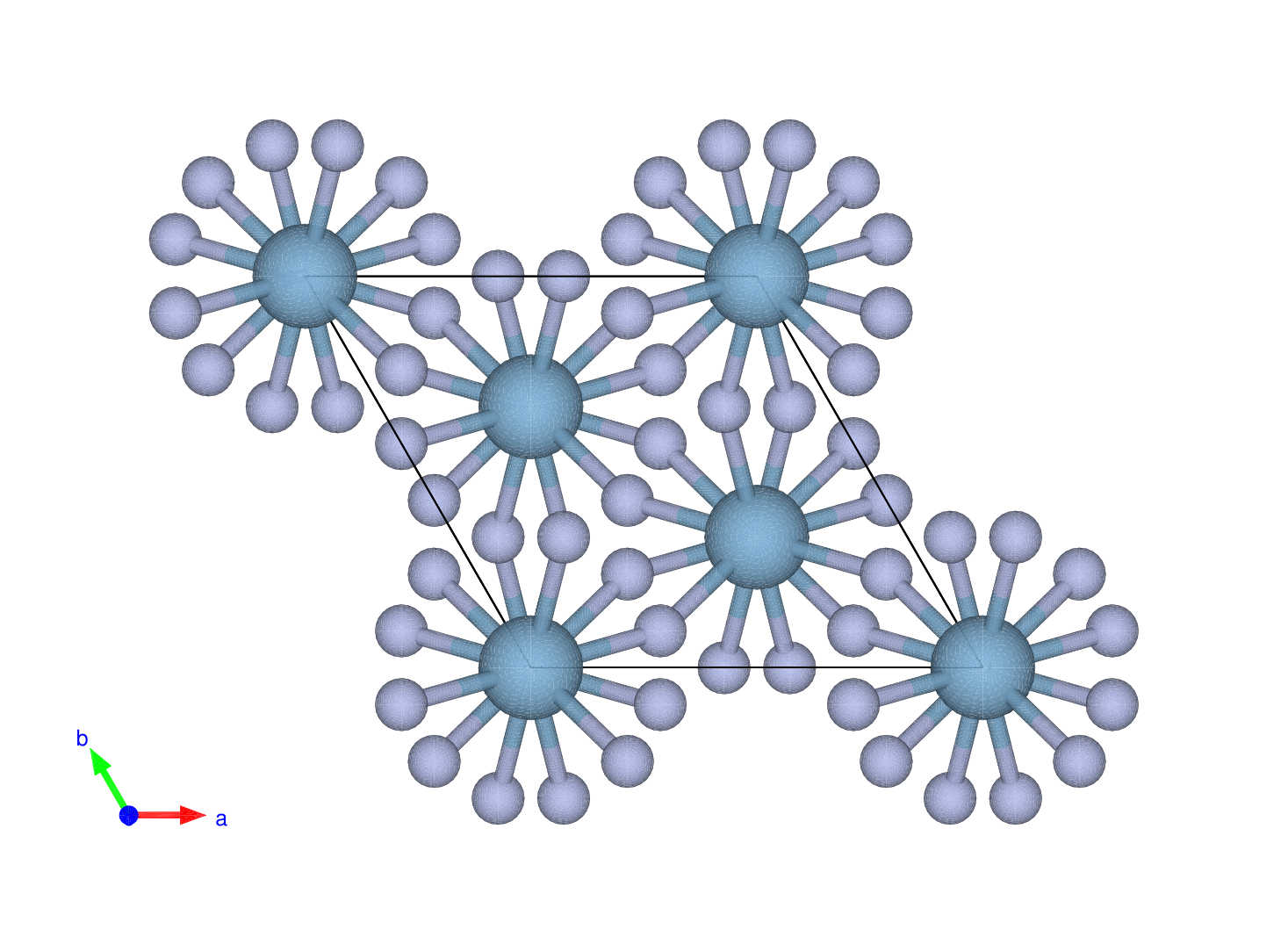}        
	\includegraphics[width=0.3\textwidth]{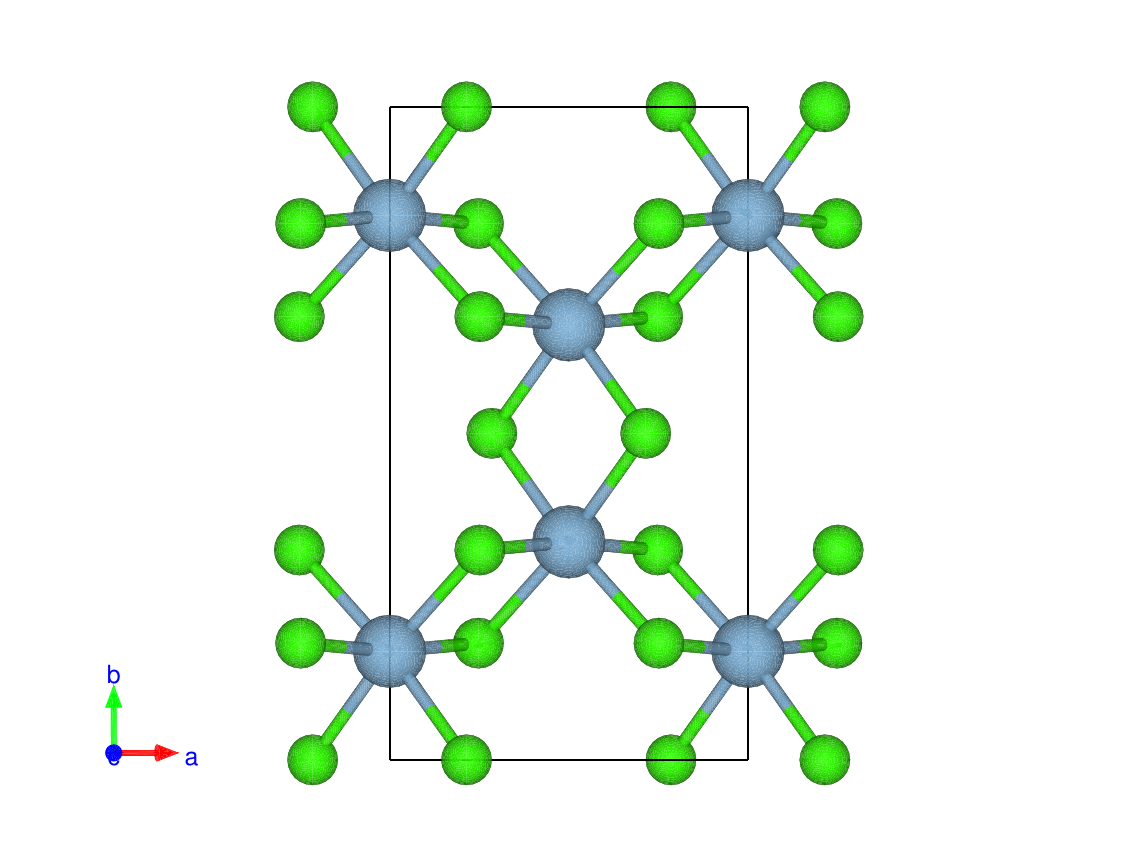}       
	\includegraphics[width=0.3\textwidth]{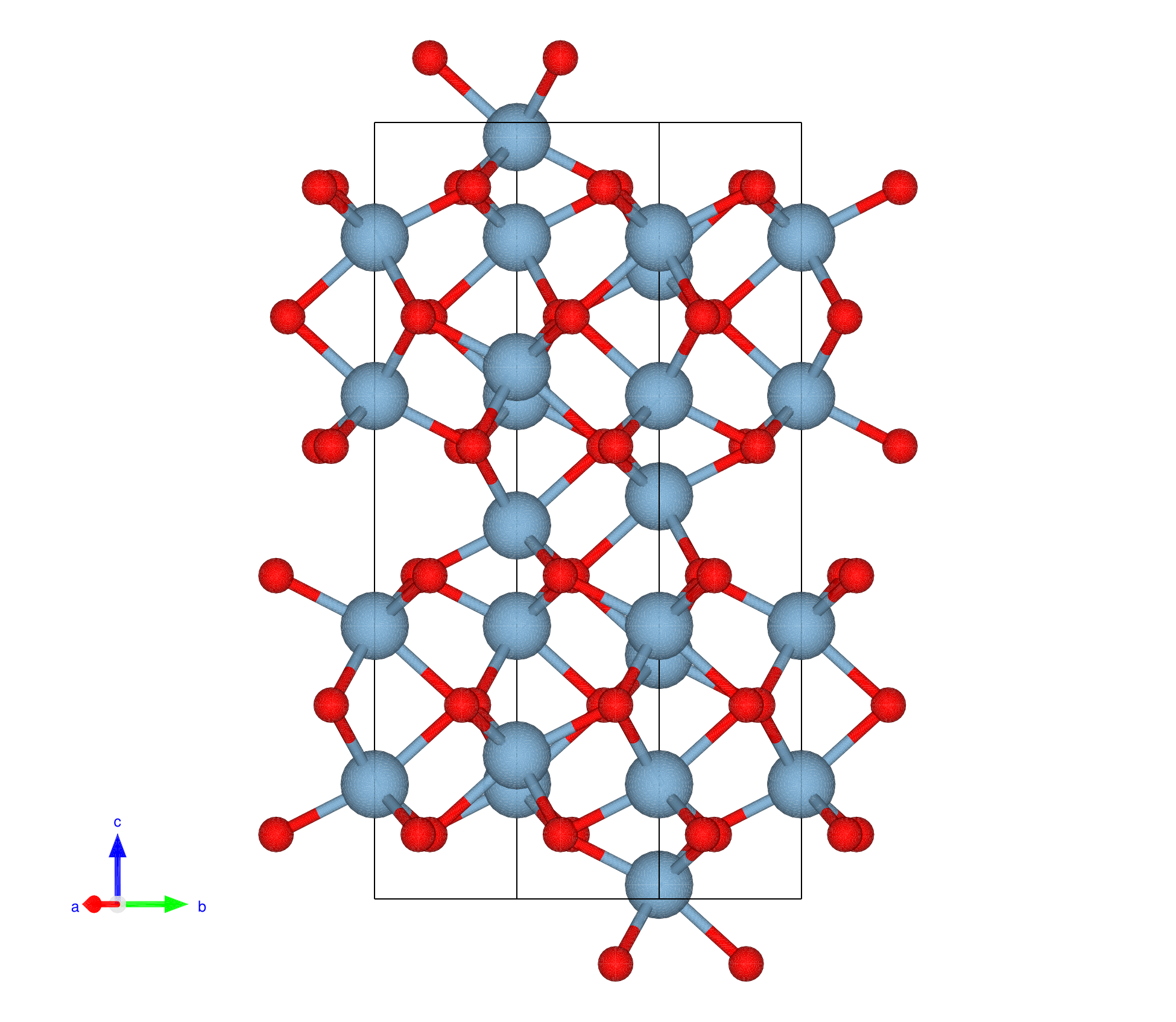}  
	\includegraphics[width=0.3\textwidth]{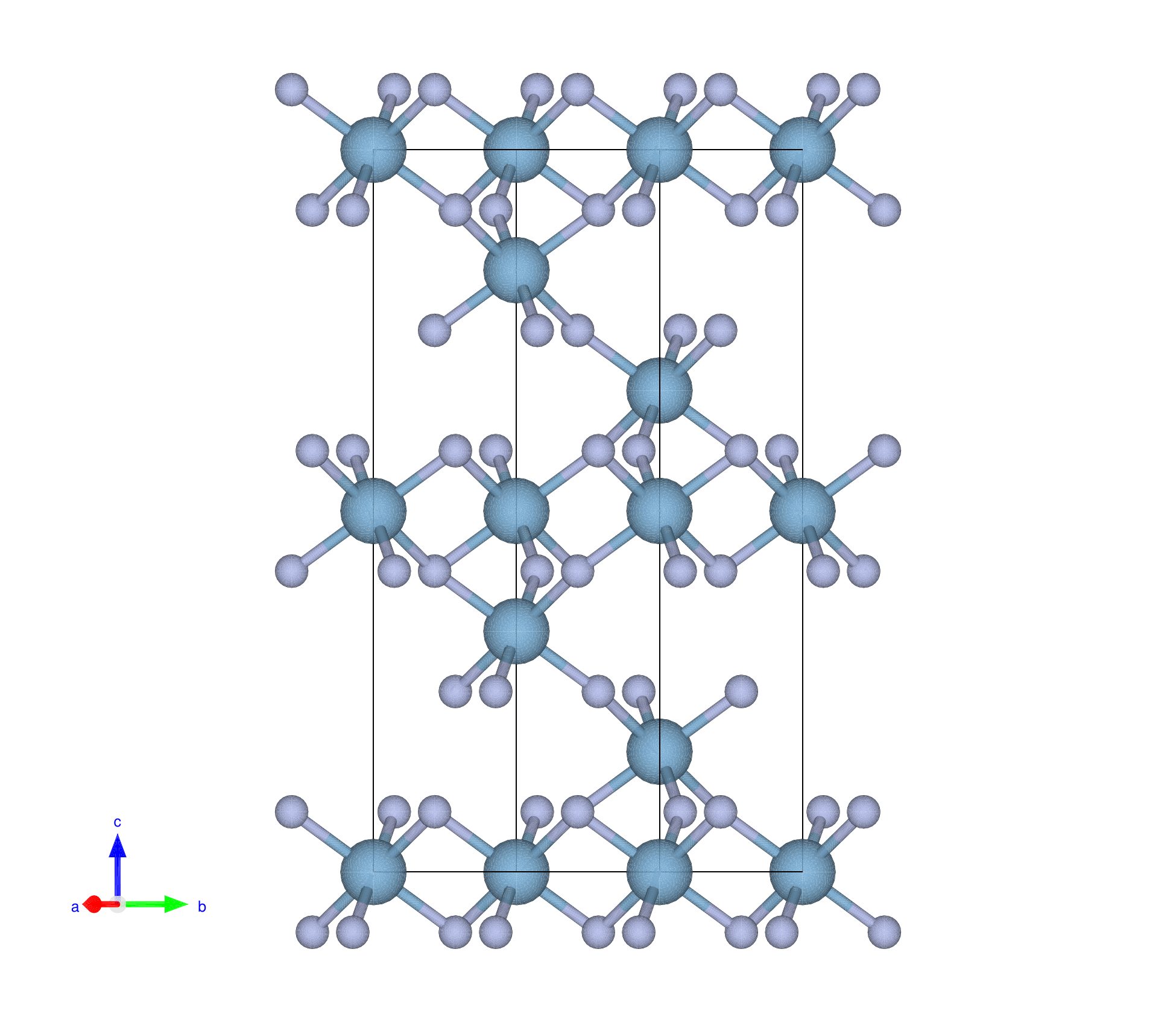}   
	\includegraphics[width=0.3\textwidth]{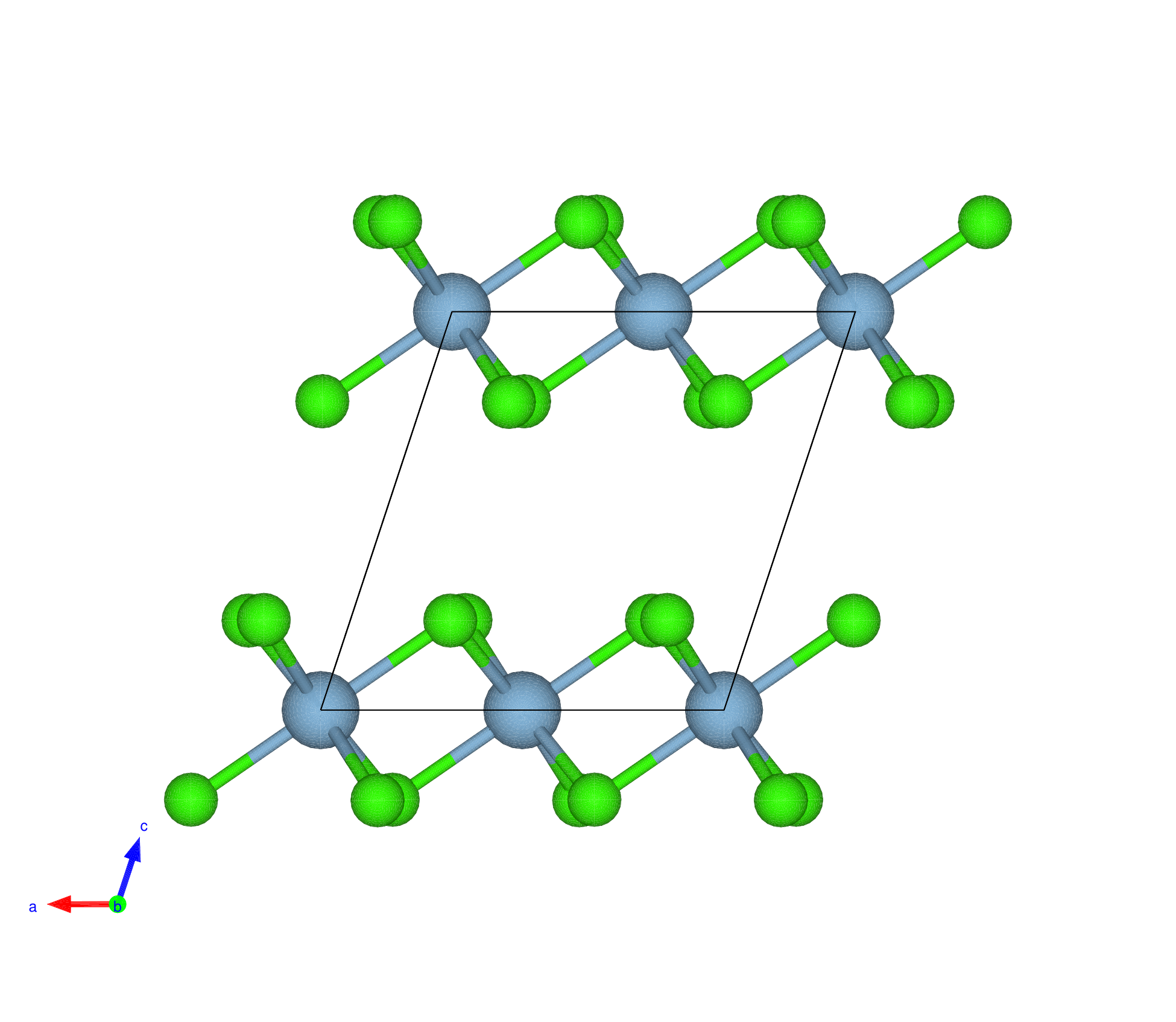}  
	\caption{Top and side views of the crystal structure of $\alpha-$Al$_2$O$_3$ (left) $\alpha-$AlF$_3$ (center) and AlCl$_3$ (right)\label{fig:structures}}
\end{figure*}

We have investigated core excitations at the aluminum K edge in three different compounds, $\alpha-$Al$_2$O$_3$, $\alpha-$AlF$_3$ and  AlCl$_3$, 
with the aim of relating their distinctive XANES features to variations in the chemical environment, i.e., the crystal structure and the ligand atom. Therefore, we will first discuss their crystallographic properties and their electronic band structures. 

Even if all the three materials contain Al in octahedral centers, their crystal structures exhibit strong differences (see Fig.~\ref{fig:structures}).
AlF$_3$ and Al$_2$O$_3$ have a rhombohedral cell (R-{3}c), with alternate layers of Al and F or O atoms in planes perpendicular to the c axis, forming octahedra linked through faces or vertices, respectively. This results in shorter distances between aluminum planes in Al$_2$O$_3$ than in AlF$_3$, even though the Al-F distances are shorter than the Al-O distances.  
On the other hand, AlCl$_3$ crystallizes in an XCl$_3$-type layer lattice with a monoclinic (C2/m) symmetry, like CrCl$_3$~\cite{Ketelaar_1947}.
In this structure, Cl-Al-Cl layers are stacked along the c axis, with octahedra sharing edges along the plane defined by the $ab$ axis, which leads to first neighbor Al atoms sharing a pair of ligands.
Finally. it should be noted that while AlF$_3$ is composed of perfectly symmetric octahedra, in Al$_2$O$_3$ and AlCl$_3$, the octahedra are slightly distorted.

\begin{figure*}[t]
	\includegraphics[width=2.0\columnwidth]{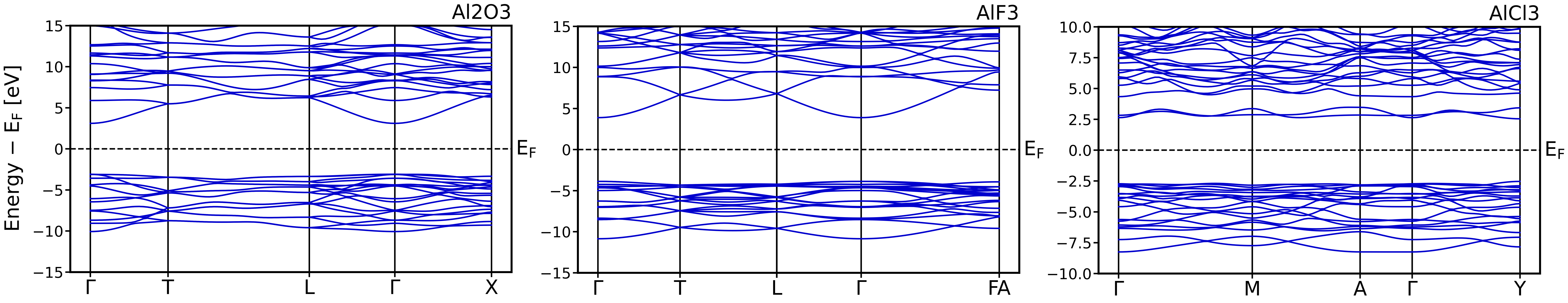} 
	\caption{Band Structure of $\alpha-$Al$_2$O$_3$ (left) $\alpha-$AlF$_3$ (center) and AlCl$_3$ (right), calculated within LDA. The zero of the energy scales has been set at the Fermi energy in the middle of the band gap.  
	\label{fig:bands}}
\end{figure*}

These strong differences reflect into the band structures, shown in Fig.~\ref{fig:bands}. 
Our LDA results for {\alo} and $\alpha$-AlF$_3$
are in very good agreement with previous KS-DFT calculations \cite{Navarro_2016, Hamed_2015, Chen_2004,Gruning_2011}. We are not aware of previous band-structure calculations for AlCl$_3$ in the literature.

The bottom conduction band of Al$_2$O$_3$ and AlF$_3$ are characterized by a typical parabolic behavior around the $\Gamma$ point, primarily composed of Al $4s$ states. The top of the valence band has contributions mostly from  $2p$ states of O or F.  
The LDA bandgap in both cases is located at the $\Gamma$ point, with values of 6.21~eV for Al$_2$O$_3$ and 7.76~eV for AlF$_3$.

The band structure of AlCl$_3$ shows more peculiar features, with weakly dispersing top-valence and two split-off bottom-conduction bands, the latter constituted by poorly hybridized Al $4s$ states, mixed with $p$ states from aluminum and chloride atoms. 
The LDA bandstructure has an indirect bandgap (between $Y$ and $\Gamma$) of 4.99~eV and a direct gap, at the $Y$ point, of 5.07~eV.

\subsection{XANES at the Al K edge \label{ssec:xanes}}

\begin{figure*}[t]
	\includegraphics[width=2.0\columnwidth]{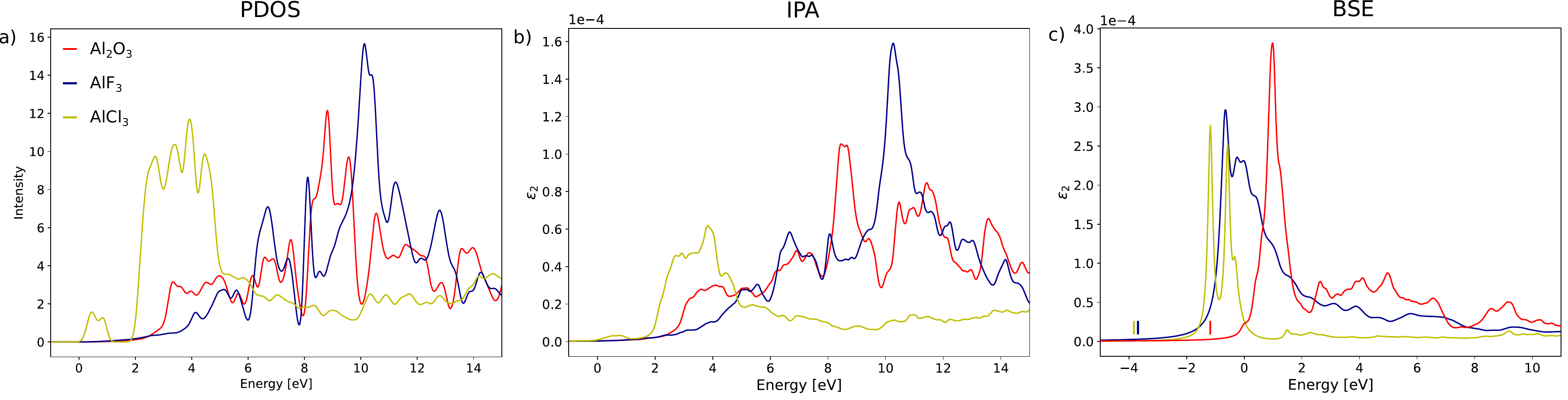} 
	\caption{(a) $p$ component of the PDOS for empty states of  aluminum atoms, where the zero of the energy axis has been set to 
	conduction band minimum energy. (b) IPA XANES spectra at the K edge of Al, where the zero has been set at the lowest transition energy of each compound. (c) Bethe-Salpeter XANES spectra at the Al K edge, setting the zero at the lowest IPA transition energy as in (b), in order to highlight the bound character of the exciton states. The IPA and BSE spectra are shown for the in-plane polarization direction, perpendicular to the $c$ axis. 
	\label{fig:xanes}}
\end{figure*}

The excitations at the Al K edge involve transitions from the $1s$ state to empty states with predominantly $p$ character, localized at the aluminum absorbing atom. Therefore, before addressing the XANES spectra, it is useful to compare first the $p$ component of the projected density of states (PDOS) on Al atoms in the three compounds, as shown in Fig.~\ref{fig:xanes}(a).
The dispersion of bottom conduction band in Al$_2$O$_3$ and AlF$_3$, shown in Fig.~\ref{fig:bands}, results in a gradual increase in 
the $p$ component of the PDOS. In contrast, the low dispersion of the two lowest conduction bands in AlCl$_3$ is reflected  as two sharp peaks in the PDOS.
It is interesting to notice that the PDOS shifts towards lower energies, as we go from AlF$_3$ through Al$_2$O$_3$ and AlCl$_3$. 
This trend corresponds to an increasing distortion of the octahedral symmetry around the Al atom, which is known to favor $sp$ hybridization~\cite{Bokhoven_2001}, and therefore, the $p$ character in the bottom of the conduction band.

The absorption spectra calculated within the IPA, which, beyond the PDOS, explicitly take into account the oscillator strenghts $\rhot_{\mu c\kvec}$, are shown Fig.~\ref{fig:xanes}(b).
In the three cases, the independent-particle spectra follow quite closely the shape of the $p$ empty PDOS of the Al absorbing atom, showing that the oscillator strenghts $\rhot_{\mu c\kvec}$ are approximately constant.
For AlF$_3$ and Al$_2$O$_3$,  the lowest-energy transitions, which are located at the zero of the energy axis of the figure, are dipole forbidden, and  therefore are not visible in any of the three polarization directions. 
On the contrary, in AlCl$_3$ there is a pre-peak, which highlights the weak $p$ character of the two non-dispersive bands at the bottom of the conduction band in Fig.~\ref{fig:bands}(c). 

The solution of the BSE gives a series of strongly bound dark and bright exciton states with energies below the lowest independent-particle transition energy. 
The absorption spectra including electron--hole interactions, shown in Fig.~\ref{fig:xanes}(c), feature pronounced excitonic peaks, followed by lower-intensity fine structures. 
These excitons have an atomic nature and a Frenkel character, associated with large binding energies. 
The comparison between IPA and BSE spectra indicates a dramatic effect of electron-hole interactions, which completely change the spectral shapes, moving their weights to lower energies. At this point, the resemblance with the PDOS is completely lost, showing that an interpretation of XANES spectra based only on a band-structure picture is inadequate.

Remarkably, there are no signatures of pre-edge features in any of the three BSE spectra as the lowest-energy excitons are dark even in AlCl$_3$.
Excitonic effects completely suppress the low intensity pre-peak features that are visible in the IPA spectrum 
in Fig.~\ref{fig:xanes}(b).
The lowest energy excitons that are dark in all the three compounds are marked with the vertical lines in Fig.~\ref{fig:xanes}(c). They have large binding energies: 1.69~eV for Al$_2$O$_3$, 4.05~eV for  AlF$_3$, and 4.33~eV for AlCl$_3$.
The calculated BSE absorption spectra of {\alo} have been compared in detail with experimental XANES in \cite{Urquiza_2024}. To the best of our knowledge, there are no accurate experimental K-edge XANES spectra for the other two compounds.
In the following sections, we will focus on AlCl$_3$, which shows  peculiar electronic structure and XANES features.

\subsubsection{Pre-peak analysis \label{item:pre-peak}}

It is well known that the K-edge XANES spectra of aluminum compounds with tetrahedral coordination have an edge position approximately 2 eV lower than those with octahedral coordination~\cite{Ildefonse_1998, Cabaret_1996}. 
However, in some cases, a visible pre-edge feature in octahedral aluminum appears at an energy similar to that of the tetrahedral compounds.
In particular, in $\alpha$-Al$_2$O$_3$ this pre-peak has been attributed to thermal vibrations that enhance octahedral distortions, allowing further $sp$ hybridization at the bottom of the conduction band~\cite{Manuel_2012, Fulton_2015}. 
Consequently, calculations including excitonic effects but without coupling to phonons fail to describe this feature, and at best, they only predict a small pre-peak at higher energies, arising from the slight distortions in corundum crystal structure~\cite{Urquiza_2024}. 
Non-resonant inelastic x-ray scattering experiments and calculations, which include excitonic effects, capture this pre-peak for high momentum transfer, confirming its non-dipolar nature~\cite{Delhommaye_2021,Urquiza_2024}.   
The origin of these dark excitations, which in the BSE spectrum result from the mixing of the independent-particle matrix elements weighted by the excitonic eigenvectors $A_{\lambda}$, are related to two factors: (1) the negligible oscillator strength of the low-energy independent-particle transitions which are dipole forbidden, (2) the cancellation of independent-particle contributions at higher energies when they are summed together~\cite{Urquiza_2023}.

In this work, for AlCl$_3$ we have found a pre-peak in the IPA spectrum, due solely to electronic contributions, see Fig.~\ref{fig:xanes}(b), which is suppressed by electron-hole interactions, as shown by Fig.~\ref{fig:xanes}(c).
This feature arises from transitions between Al $1s$ and the bottom of the conduction band with a weak $p$-character, shown by the PDOS in Fig.~\ref{fig:xanes}(a). 
The origin of the pre-peak is the breaking of the octahedral symmetry around the Al atoms, which allows for $sp$ hybridization.  
However, when the independent-particle transitions mix together in Eq.~(\ref{eq:spectrumBSE}) to give the BSE spectrum, two prominent peaks and a shoulder emerge at the onset, while suppressing any signature of a pre-peak below them. 
This raises two important questions: Why is the pre-peak absent in the BSE spectra? What is the nature of the lowest-energy dark and bright  excitons?

\begin{figure}
	\includegraphics[width=0.75\columnwidth]{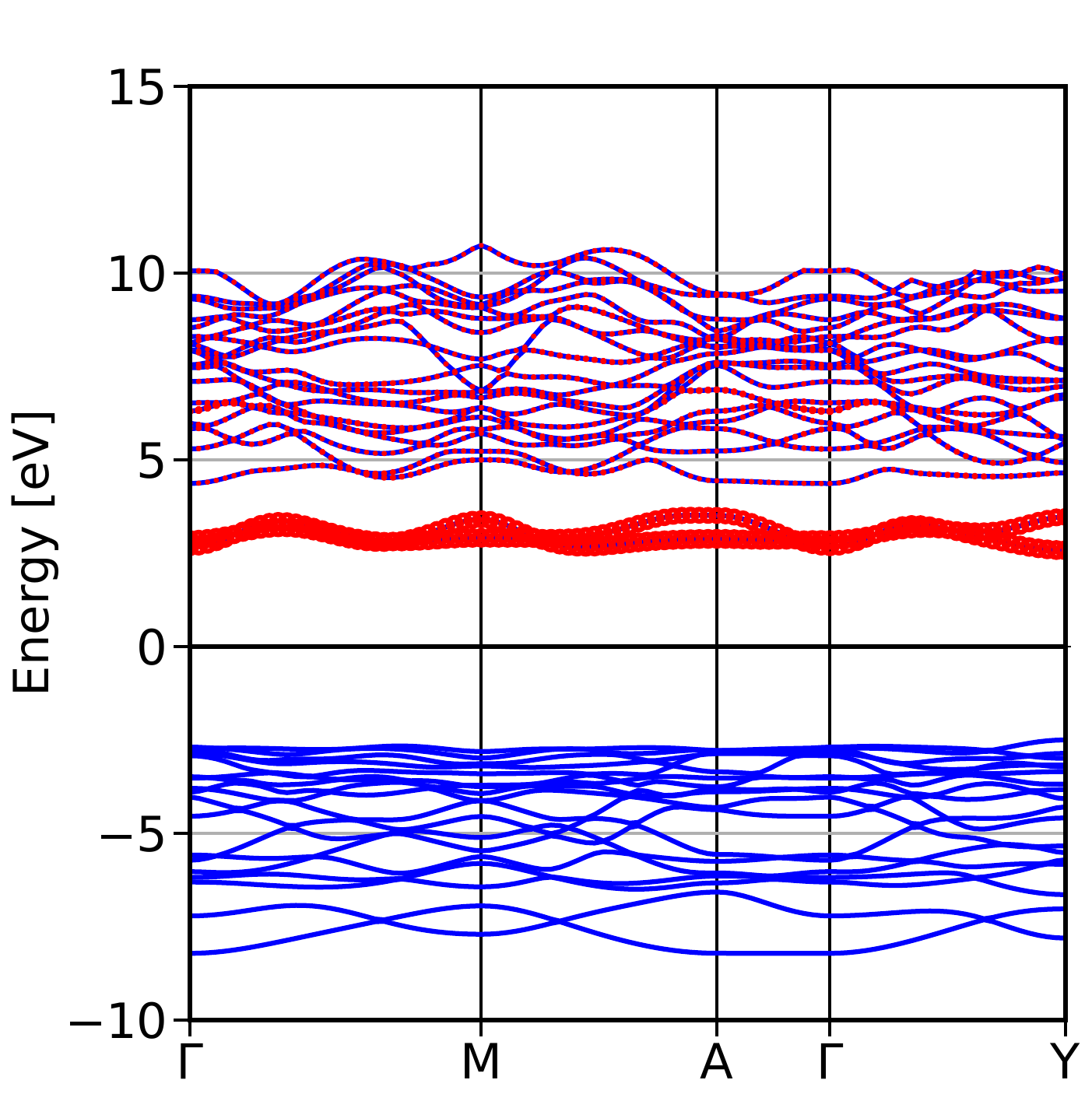}  
	\includegraphics[width=0.75\columnwidth]{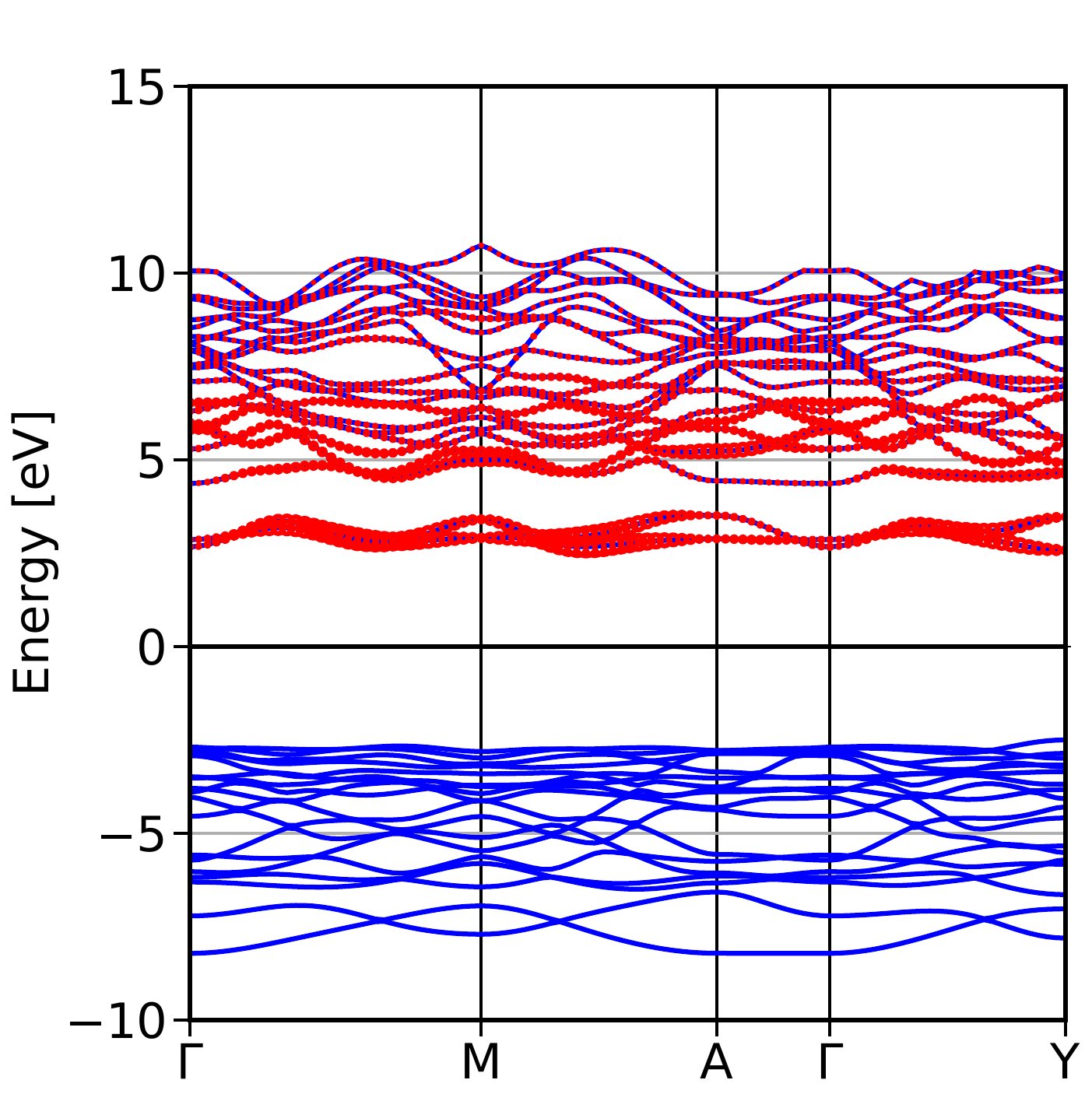}  
	\caption{Contributions of independent transitions to the lowest-energy dark (top) and bright (bottom) excitons. The size of the circles is proportional to the excitonic coefficients $ | A_{\lambda}^{vc{\bfk}}| $.\label{fig:exciton}}
\end{figure}

Fig.~\ref{fig:exciton} illustrates the excitonic weight $| A^{\mu c\kvec}_\lambda |$, projected on the LDA band structure, of each independent particle transition $\mu\kvec \rightarrow  c\kvec$ contributing to the first dark ($\lambda = 1$) and bright ($\lambda = 5$) excitons. Each independent-particle transition $\mu\kvec \rightarrow  c\kvec$ is then represented by a circle in the conduction band, with a size proportional to their weight.
In both excitons, the most important contributions are given by transitions from core $1s$ states (out of view in this scale) 
to the  non-dispersive bottom conduction bands. 
The dark exciton is composed exclusively by these 
transitions - see top panel - while for the bright exciton  transitions from Al $1s$ to upper conduction bands also sum up - see bottom panel. Moreover, the strong delocalization in reciprocal space of the  excitonic weights $| A^{\mu c\kvec}_\lambda |$ suggests a high localization of the exciton in real space, consistent with the typical Frenkel-model picture.
From Fig.~\ref{fig:xanes}(b), we know that the oscillator strenghts, $\rhot_{\mu c\kvec}$, for those transitions are dipole allowed. 
Therefore, each contribution $A^{\mu c\kvec}_\lambda \rhot_{\mu c\kvec}$  is separately not zero in Eq.~(\ref{eq:spectrumBSE}). However, the different contributions cancel out when they are summed together and give a dark exciton. 
This picture contrasts with other materials, including Al$_2$O$_3$, where the strongly bound dark excitons arise instead from zero oscillator strengths $\rhot_{\mu c\kvec}$.

\subsubsection{Anisotropy}

\begin{figure*}[htbp]
	\includegraphics[width=2\columnwidth]{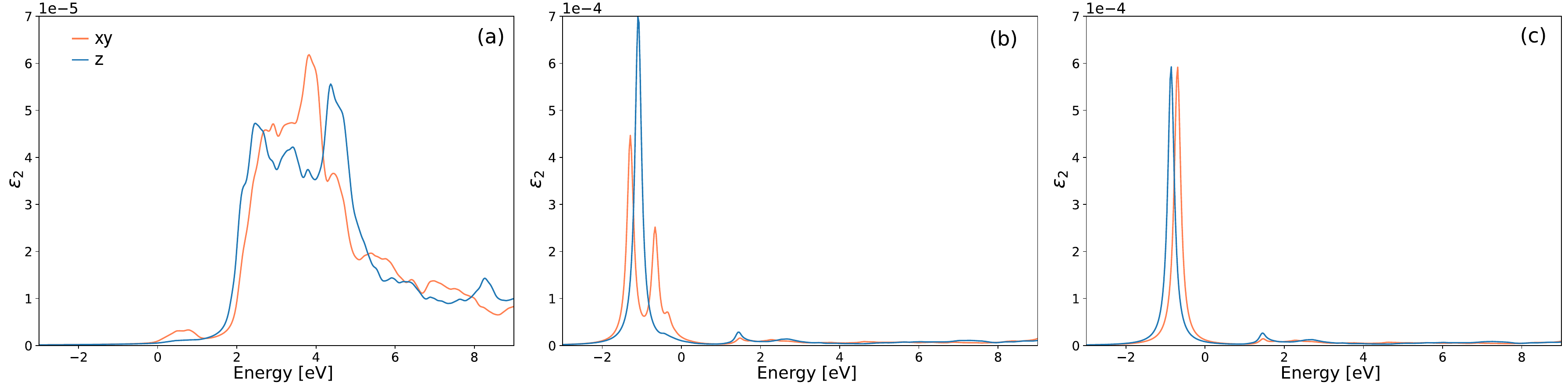} 
	\caption{ XANES spectra at the Al K edge in AlCl$_3$ along two polarization directions $xy$ (perpendicular to the $c$ axis) and $z$ (parallel to the $c$ axies), calculated with (a) the independent particle approximation, (b) the Bethe-Salpeter equation, and (c) with the BSE, but excluding explicitly the two lowest-energy unoccupied bands.\label{fig:xanes_alcl3}}
\end{figure*}

The anisotropy in the absorption spectra of Al$_2$O$_3$ has been extensively studied for 
core~\cite{Cabaret_2009, Manuel_2012,Urquiza_2024} and optical spectra~\cite{Marinopoulos_2011,Urquiza_2023}, whereas 
AlF$_3$ does not exhibit significant directional differences in absorption spectra~\cite{Chen_2004}.
Therefore, in the present section we will focus on the more interesting case of AlCl$_3$.

AlCl$_3$ has a strongly anisotropic crystal structure associated to its layered configuration. As a result, the XANES spectra along the polarization directions $xy$ (perpendicular to the $c$ axis) and $z$ (parallel to the $c$ axis) are quite different, as shown in Fig.~\ref{fig:xanes_alcl3} for (a) IPA and (b) BSE. 
In the IPA spectrum, the pre-peak is completely suppressed in the $z$ direction, while at higher energies the spectrum behaves similarly along the two directions. 
In the BSE spectrum, the two bright excitons in the $xy$ direction, which are split by 0.62~eV and give rise to the two main features, are dark in the $z$ direction. Instead, a single bright exciton is responsible for the main feature in the $z$ polarization, appearing about 0.2 eV higher than the edge in the $xy$ polarization.
The origin of this anisotropy is related to the contributions of the first two unoccupied bands to  the excitonic peaks. 
In fact, BSE calculations that explicitly exclude these two empty bands, as shown in Fig.~\ref{fig:xanes_alcl3}(c), yield a series of dark and bright excitons where the anisotropy is strongly reduced, resulting in only a small shift towards lower energies in the $z$ direction.

\section{Conclusions \label{sec:conclusions}}

In conclusion, we have investigated the XANES spectra of a series of aluminum compounds with octahedral coordination to evaluate the influence of chemical environments on their spectral features. 
By solving the Bethe-Salpeter equation, to account for excitonic effects in the XANES spectra, we found that the three materials exhibit a strong electron-hole interaction,  
leading to the formation of localized Frenkel-like excitons with large binding energies.
Our analysis reveals that octahedral distortion can be a primary mechanism that allows for $sp$ hybridization, enabling atomic-like dipole transitions $s\rightarrow p$, which are  the main contribution to these excitons.
Despite this, and in line with other octahedral compounds, the first excitons in all three materials are dark, resulting in the absence of pre-edge features in the calculated XANES spectra. 
We analyzed in detail the case of AlCl$_3$, where two non-dispersive bands at the bottom of the conduction band lead to excitons with noteworthy properties. 

Our calculations demonstrate that an atomic perspective alone has strong limitations to analyze the XANES spectra of aluminum in different chemical environments, as the ligand states strongly hybridize with the Al $4s$ orbitals, significantly influencing the electronic structure. In addition, excitonic effects play a crucial role in determining the edge position and absorption features. 
As a result, establishing a direct correlation between the differences in the XANES spectra of the three compounds and changes in the chemical environment, through modifications of the ligand atom and variations in the crystal structure, remains challenging.

\begin{acknowledgments}
We acknowledge the African School on Electronic Structure: Methods and Applications (ASESMA) where this work was initiated. NA and AA also thank the Palaiseau Theoretical Spectroscopy Group for the kind hospitality through a research stay that was possible thanks to the ``Dispositif de Soutien aux Collaborations avec l’Afrique subsaharienne'' of the CNRS, supported also by the Ecole Polytechnique.
We acknowledge the French Agence Nationale de la Recherche (ANR) for financial support (Grant Agreements No.  ANR-19-CE30-0011). Computational time was granted by GENCI (Project  No.  544).   
\end{acknowledgments}

\bibliography{bib}

\end{document}